# Algorithms For Longest Chains In Pseudo-Transitive Graphs

Farhad Shahrokhi

Department of Computer Science and Engineering, UNT
P.O. Box 13886, Denton, TX 76203-3886, USA
farhad@cse.unt.edu

**Abstract**

A directed acyclic graph $G = (V, E)$ is pseudo-transitive with respect to a given subset of edges $E_1$, if $ab \in E_1$ and $bc \in E$ implies that $ac \in E$. We give algorithms for computing longest chains and demonstrate geometric applications that unify and improves some important past results.

## 1 Introduction and Summary

The notation of an "ordering" naturally arises in many problems of combinatorial and computational geometry [8] and the underlying structures usually exhibit some sorts of "weak transitivity" properties. However, in many cases the crucial properties of these problem can not be categorized by the classical theory of partially ordered sets [5,9]. Motivated by geometric problems we introduce here the concept of "pseudo-transitivity" in the directed acyclic graphs. We derive algorithms for computing the longest chains under certain pseudo-transitivity assumptions, and demonstrate their applications in computing the maximum independent sets in the intersection graph of geometric objects.

The remaining of this section contains crucial definitions and the descriptions of the results. Throughout this paper, $G=(V, E)$ is a directed acyclic graph with $|V|=n$ and $|E|=m$. A chain in $G$ is a directed path $C: a_1, a_2, ..., a_k$, $k \geq 1$ so that for any $1 \leq i < j \leq k$, $a_i a_j \in E$. It is important to note that there is a one-to-one correspondence between the chains in $G$ and the cliques in the undirected graph obtained from $G$ by ignoring the direction of the edges, since $G$ is acyclic. An anti-chain in $G$ is an independent set of vertices in $G$. Let $\omega(G)$ and $\alpha(G)$ denote the largest number of vertices in any chains and in any antichains of $G$, respectively. A set C of chains (antichain) covers $V$, if every vertex in $V$ is located on exactly one chain (antichain) in C. Let $\chi(G)$ and $\beta(G)$, denote respectively, the minimum number of antichains and the minimum number of chains that cover $V$. Observe that $\chi(G)$ and $\beta(G)$ are the chromatic number and the clique cover number, respectively, of the undirected graph underlying $G$.

If $E$ is partial order on $V$, that is if $ab \in E$ and $bc \in E$ implies that $ac \in E$, then $G$ is called a *transitive* graph or a *Poset*. Dilworth's theorem [5] asserts that for any Poset $G = (V, E)$, $\chi(G) = \omega(G)$ and $\alpha(G) = \beta(G)$. Moreover, in this case low order polynomial time algorithms have been known for the problems of computing $\chi(G)$, $\omega(G)$, $\alpha(G)$ and $\beta(G)$, which in general are $NP$-hard [7].

Let $G = (V, E)$ and $E_1 \subseteq E$ with the property that for any $a, b, c \in V$, $ab \in E_1$ and $bc \in E$ implies that $ac \in E$. Then we say $G$ is *pseudo-transitive* with respect to $E_1$, and write $G = (V, E_1, E)$. When the context is clear we say $G=(V, E_1, E)$ is pseudo-transitive. Note that by selecting $E_1=\emptyset$, any directed acyclic graph $G$ becomes pseudo-transitive. However in this extreme case pseudo-transitivity becomes a useless property. Note further that if $E_1 = E$, for a pseudo transitive graph $G = (V, E_1, E)$, then $G = (V, E)$ is a transitive graph, that is, $G$ is a Poset. To get a feeling about the geometric aspects of pseudo-transitivity consider the following example. Let P be a finite collection of bounded closed subsets of $R^k$. Let $P, Q \in$ P. We say that $Q$ and $P$ are *disjoint*, if $P \cap Q = \emptyset$, otherwise we say $P$ and $Q$ intersect. The *intersection graph* of P is an undirected graph with the vertex set P whose edges are intersecting (un-ordered) pairs of elements of P. Let $h$ be a $k$-dimensional hyperplane in $R^k$. We will construct a pseudo-transitive graph $G(P,h)$ whose underling undirected graph is isomorphic to the complement of the intersection graph of P. With no loss of generality assume that the $h$ is perpendicular to the $x$ axis, if this is not the case, we can rotate the coordinate system so that this holds. Define the *left most* vertex of $P \in$ P to be a vertex with the smallest $x$ coordinate. We say that $Q$ is *disjoint from P*, if $P$ and $Q$ are disjoint and the $x$ coordinate of the leftmost vertex of $P$ is smaller or equal to the $x$ coordinate of a leftmost vertex of $Q$. Let $P, Q \in$ P so that $Q$ is disjoint from $P$. We say that $P$ is *to the left* of $Q$, if there is hyperplane parallel to $h$ that separates $P$ from $Q$ in $R^k$ so that $P$ is located to the left of $h$, and $Q$ is to the right of $h$, where right and left correspond to the positive and negative directions of the $x$ axis, respectively.

**Theorem 1.1** *There is directed pseudo-transitive graph $G_{(P,h)} = (P, E_1, E)$ whose underlying undirected graph is isomorphic to the complement of the intersection graph of* P.

Proof. Define $E_1$ to be the set of all edges of form $PQ$, $P, Q \in$ P so that $P$ is to the left of $Q$. To construct $E$, consider the following simple algorithm. Let initially $E = E_1$, and consider all disjoint pairs $P, Q$ so that $Q$ is disjoint from $P$, $PQ \notin E$, and the $x$ coordinate of the leftmost vertex of $P$ is smaller that the $x$ coordinate of leftmost vertex of $Q$. We include any such $PQ$ in $E$. To finish the construction of $E$, consider all remaining disjoint pairs $P, Q$ for which the $x$ coordinate of the leftmost vertex of $P$ equals to the $x$ coordinate of leftmost vertex of $Q$. For any such pair add $PQ$ to $E$, if $E \cup \{PQ\}$ is acyclic, otherwise add $QP$ to $E$. It is easy to verify that $G_{P,h}$ is acyclic, and that $PQ \in E_1$ and $QR \in E$ implies that $PR \in E_1$. Moreover, $PQ \in E$ iff the $P$ and $Q$ are disjoint, and hence $G_{(P,h)}$ is isomorphic to the complement of the intersection graph of P. ∎

Let $G = (V, E_1, E)$ be pseudo-transitive and denote by $G_1$ and $G_2$ the directed subgraphs of $G$ with the vertex sets $V$, and the edges sets $E_1$ and $E_2 = E - E_1$, respectively. Furthermore, let $\omega(G)$ and $\omega(G_2)$ denote number of vertices in the longest chains of $G$ and $G_2$, respectively. In section two we show that for any pseudo-transitive $G = (V, E_1, E)$, $\omega(G)$ can be computed in $O(\omega(G_2)n^{\omega(G_2)+2})$ time. The algorithm has nice applications in graph drawing [4], and when applied to a collection on unit height axis parallel rectangles, its running time is slightly better than the original Dynamic Programing Algorithm of Agarwal, Kreveld, and, Suri [3]. Pach and Töröcsik [8], and others [10] studied extremal problems for disjoint line segments in the plane. They discovered that the "disjointness properties" for the line segments can be captured by the union of several partial orders that all act on the line segments. The work in [8] inspired us to study the pseudo-transitivity when both $E_1$ and $E - E_1$ are partial relations on $V$. In Particular we have shown in Theorem 2.3 that in this case $\omega(G)$ can be computed in $O(nm)$ time. The result has interesting computational consequences when applied to well known problems relating to partial orders. See for instance, the work of Biro and Trotter on segments orders [2]. Specifically, we can compute a largest set of disjoint line segments in a set *S* of *n* line segments that all have one end point on a common line *l* and make an acute angel with *l*, in $O(n^3)$ time. If the segments in *S* have one end point on *l* but make arbitrary angels with it, then we can compute 1/2-optimal solution in $O(n^3)$ time. Moreover, the algorithm also can be used for circle graphs [6] and gives an optimal disjoint (independent) set in $O(n^3)$ time. Agarwal and Mustafa [1] have derived approximation and exact algorithms for maximum independent sets in much more general versions of planar line segments as well as convex objects. However, their algorithms when applied to the cases, discussed here, give weaker results.

## 2  Algorithms

We will view chains as sets of vertices. For a chain *C* in $G = (V, E)$, let $|C|$ denote the length or the number of vertices in *C*. Let $C: a_1, a_2, ..., a_k$ and $C': b_1, b_2..., b_t$ be vertex disjoint chains in *G*. We denote the ordered set $\{a_1, a_2,...,a_k, b_1, b_2,...,b_t\}$ by $C \cup C'$.

For a chain $C: a_1, a_2,...,a_k$, let $\hat{\omega}(C)$ denote the maximum number of vertices in any chains of the form $a_1, a_2, ..., a_k, ..., a_t$, where $t \geq k$.

Let *C* be a chain in *G* with $|C| \geq \omega(G_2) + 1$. Note that *C* must contain at least two vertices *a* and *b* so that $ab \in E_1$. We refer to *a* as a *pivoting* vertex in *C*.

**Lemma 2.1** *Let $G = (V, E_1, E)$ be pseudo-transitive. Let C be a chain in G with $|C| \geq \omega(G_2) + 1$ and let a be a pivoting vertex in C, then the following holds.*

*(i) Let C' be a chain, then $C \cup C'$ is a chain if and only if $C - \{a\} \cup C'$ is a chain.*

*(ii)* $\hat{\omega}(C) = \begin{cases} Max_{a^* \in V}\, \hat{\omega}\big(\{C - \{a\}\} \cup \{a^*\}\big) + 1 & \text{there is a vertex } a^* \text{ so that } C \cup \{a^*\} \text{ is a chain} \\ |C| & \text{Otherwise} \end{cases}$

Proof. For (*i*) it suffices to show that if $\{C - \{a\}\} \cup C'$ is a chain so is $C \cup C'$. To see this note that there is a vertex $b \in C - \{a\}$ so that $ab \in E_1$, and note that $bc \in E$ for any $c \in C'$, since $\{C-\{a\}\} \cup C'$ is a chain. It follows that from pseudo-transitivity that $ac \in E$ for any $c \in C'$, verifying the claim in (i). For (ii), when $C$ can not be extended to a larger chain, then clearly, $\hat{\omega}(C) = |C|$. To complete the proof of (ii) note that $\hat{\omega}(C) = \max_{a^* \in V} \hat{\omega}(C \cup \{a^*\})$, where the maximum is taken over all vertices $a^*$ so that $C \cup \{a^*\}$ is a chain. However, by Part (i), $C \cup \{a^*\}$ is a chain if and only if $\{C - \{a\}\} \cup \{a^*\}$ is a chain, completing the proof. ∎

For any vertex $x \in V$, let $N(x)$ denote the set of all vertices adjacent from $x$, where we assume that $x \in N(x)$. Let $C$ be a chain in $G$ with $|C| = \omega(G_2) + 1$ and let $a$ be a pivoting vertex on $C$. Let $a^* \in N(a)$ so that $C' : \{C - \{a\}\} \cup \{a^*\}$ is a chain, then we say $C$ generates $C'$.

Let $G = (V, E_1, E)$ be pseudo-transitive, define the *transition graph of $G$* to be a directed graph $G^\omega = (V^\omega, E^\omega)$ with $V^\omega$ being the set of all chains $C$ in $G$ with $|C| = \omega(G_2) + 1$ so that $CC' \in E^\omega$ iff $C$ generates $C'$. Note that $G^\omega$ is acyclic. The following observation is a consequence of Part (ii) in the preceding Lemma.

**Observation 2.1** Let $G = (V, E_1, E)$ be pseudo-transitive and $\omega(G) \geq \omega(G_2) + 1$. Then $\omega(G)$ equals to the length of a longest path in $G^\omega$ plus $\omega(G_2) + 1$. n

**Theorem 2.1** Let $G = (V, E_1, E)$ be pseudo-transitive. Then $\omega(G)$ can be computed in $O(n^{\omega(G_2)}m)$ time.

Proof. We store $G$ in the adjacency matrix form and also in the adjacency list form. The algorithm has three stages. In stage one, we compute $\omega(G_2)$, using a brute force method in $O\left(\sum_{r=1}^{\omega(G_2)+1} \binom{n}{r} r^2\right)$ time which is $O(\omega(G_2) n^{\omega(G_2)+1})$ time, using $O(n^2)$ storage. (Note that it takes $O(r^2)$ time to check if an $r$-subset is a chain.) In the second stage we check to see if $G$ has a chain of $\omega(G_2)+1$ vertices in $O(n^{\omega(G_2)+1})$ time, and $O(n^2)$ storage. If there is a chain of $\omega(G_2) + 1$ vertices in $G$, then the algorithm starts stage three, otherwise it stops and reports that $\omega(G) = \omega(G_2)$. At the stage three of the algorithm we compute a longest path in the transition graph $G^\omega$ using a topological ordering of vertices in $V^\omega$, and output the number of vertices in such a path plus $\omega(G_2)+1$. To construct $V^\omega$ we construct all $\omega(G_2)+1$-subsets of $V$ and check in $O((\omega(G_2)+1)^2)$ time if they are chains. Once such a chain $C$ is generated, all vertices adjacent to $C$ are also generated and stored in the adjacency list format. Let $C$ be a chain in $G^\omega$ starting at $x \in V$, then degree of $C$ in $G^\omega$ is at most $deg(x)$. Thus, once $C$ is constructed it takes $O(deg(x)\omega(G_2))$ to construct all chains $C'$ generated by $C$ and store them. It follows that adjacency list of the transition graph can be constructed in

$$O\left(\binom{n}{\omega(G_2)+1}(\omega(G_2)^2)\right) + O\left(\sum_{x \in V}\binom{n-1}{\omega(G_2)}deg(x)\omega(G_2)\right)$$

or $O(n^{\omega(G_2)}m)$ time and storage. This finishes the proof, since topological ordering of any acyclic graph can be done in a linear time of the input length. ∎

Let $G = (V, E_1, E)$ be pseudo-transitive so that both $E_1$ and $E - E_1$ are partial orders on $V$, then we call $G$ *strongly pseudo-transitive*. Note that in this case $ab \in E_1$ and $bc \in E_1$ implies that $ac \in E_1$, $ab \in E - E_1$ and $bc \in E - E_1$ implies that $ac \in E - E_1$, and finally $ab \in E_1$ and $bc \in E - E_1$ implies that $ac \in E$. Our next result is that in a strongly pseudo-transitive graph a maximum weighted chain can be computed in low order polynomial time. For any $x \in V$, let $c_x$ denote the weight or the cost of vertex $x$.

Let $G = (V, E_1, E)$ be strongly pseudo-transitive and let $C: a_1, a_2, ...,a_k$, $k \geq 3$ be a chain. We say that $C$ is *splitable*, if there is a $j$, $2 \leq j \leq k-1$ so that for $i = 1, 2, ..., j-1$, $a_i a_j \in E_1$. In this case $a_j$ is called a *splitting element*. We say $C$ is *degenerate* if it is not splitable.

Lemma 2.2 *Let $G = (V, E_1, E)$ be strongly pseudo-transitive, then, the following hold.*

*(i)- Let $C_1$: $a_1, a_2, ...,a_t$ and $C_2$: $a_t, a_{t+1}, ..., a_p$ be chains in G, so that $a_i a_t \in E_1$ for $i = 1, 2, 3, ..., t-1$. Then, $C_1 \cup C_2$: $a_1, a_2, ...,a_p$ is a chain in G.*

*(ii)- Let $C$: $a_1, a_2, ...,a_p$, $p \geq 3$ be a chain in G. Then C is degenerate if and only if for $i = 2, ..., p-1$, $a_1 a_i \in E - E_1$.*

*(iii)- Let $C = a_1, a_2, ..., a_p$, $p \geq 3$ be a chain in G. If C is splitable, then there is an splitting element $a_j$, $1 < j < p$ such that the chain $C' = a_j, a_{j+1}, ..., a_p$ is degenerate.*

Proof. To prove (i), let $a_i \in C_1$ and $a_j \in C_2$, $i, j \neq t$. We must show that $a_i a_j \in E$. Clearly, $a_i a_t \in E_1$ and $a_t a_j \in E$, then $a_i a_j \in E$, proving the claim. For (ii), clearly, if for $i=2, 3,..., p-1$, $a_1 a_i \in E - E_1$, then $C$ is degenerate. For converse note that for $p = 3$ the claim holds, and observe that for $p \geq 4$, if $C$ is degenerate, then $a_1, a_2, ..., a_{p-1}$ is also degenerate and proceed by induction on $p$. For (iii) let $j$ be the largest integer so that $aj$ is a splitting element for $C$, and let $C'$ be $a_{j+1}, a_{j+2}, ..., a_k$. Assume to the contrary that $C'$ is splitable, and let $a_t, t > j$ be a splitting element for $C'$, then $a_t$ is also a splitting element for $C$ contradicting the maximality of $j$. ∎

Remarks. Part (i) of the Lemma shows how to construct long chains from smaller ones. Part (ii) explores the structure of the degenerate cases.

We will need the following crucial terms and definitions, which will be used to design our dynamic programming algorithm. For any $xy \in E, x, y \in V$, let $\omega_{x,y}$ denote the cost of a maximum cost chain starting at $x$ and terminating at $y$ in $G$. Moreover, for any $xy \in E_1, x, y \in V$, let $\bar{\omega}^1_{x,y}$ denote the cost of a maximum cost chain $C$ from $x$ to $y$ so that for any element of $z \neq y$ of $C$, it holds that $zy \in E_1$. Note that the cost of a maximum cost chain in $G$ is the Maximum of all $\omega_{x,y}$.

Let $xy \in E$, $x, y \in V$, and let $\bar{\omega}_{x,y}$ denote the maximum cost of a degenerate chain among those degenerate chains starting at $x$, and finishing at $y$. Finally, let $x, y \in X$ so that $xy \in E$, and let $\bar{\omega}^1_{x,y}$ denote the maximum cost of a degenerate chain among those degenerate chains starting at x and y, so that for any element $k \neq x, y$ of the chain, $ky \in E_1$.

Theorem 2.2 *let $G = (V, E)$ be strongly pseudo-transitive, then, the following hold.*

(i)
$$\omega_{x,y} = max \left\{ \begin{matrix} max \\ t \\ xt \in E_1 \\ ty \in E \end{matrix} \{\omega^1_{x,t} + \omega_{t,y} - C_t\}, \bar{\omega}_{x,y} \right\}$$

(ii)

$$\omega^1_{x,y} = max \left\{ \max_{\substack{xt \in E_1 \\ ty \in E_1}} \{\omega^1_{x,t} + \omega^1_{t,y} - C_t\}, \overline{\omega}^1_{x,y} \right\}$$

(iii)

$$\overline{\omega}_{x,y} = max \left\{ \max_{\substack{xt \in E-E_1 \\ ty \in E}} \{\overline{\omega}^1_{x,t} + \overline{\omega}_{t,y} - C_t\}, \max_{\substack{xt \in E-E_1 \\ ty \in E}} \{\overline{\omega}^1_{t,y} + c \} \right\}$$

(iv)

$$\overline{\omega}^1_{x,y} = max \left\{ \max_{\substack{xt \in E-E_1 \\ ty \in E_1}} \{\overline{\omega}^1_{x,t} + \overline{\omega}^1_{t,y} - C_t\}, \max_{\substack{xt \in E-E_1 \\ ty \in E_1}} \{\overline{\omega}^1_{t,y} + c_x \} \right\}$$

Proof. We prove the claim using Lemma 2.2 and distinguishing between the splitable and degenerate cases. Let $C$ be the desirable chain. In (i) and (ii), the first expression inside of the first inner curly bracket applies when $C$ is splitable, whereas the remaining expression applies when $C$ is degenerate. For (iii) and (iv), the expression inside of the first inner curly bracket applies when $C - \{x\}$ is separable and the expression inside of the second curly bracket applies when $C - \{x\}$ is degenerate. In particular, for (iii) if $C - \{x\}$ is splitable, Part (iii) of Lemma 2.2 applies and identifies, a splitting element $t$, and subchains $C_1$ and $C_2$ of $C$, $C - \{x\} = C_1 \cup C_2$ so that $C_2$ is degenerate. Note that $xt \in E - E_1$, since $C$ is degenerate. Moreover, $C_1 \cup \{x\}$ must be degenerate, since otherwise $C$ is splitable. Now assume for (iii) that $C-\{x\}$ is degenerate, and let $t$ be the starting vertex of $C-\{x\}$ and observe that $xt \in E - E_1$. Details of (iv) are similar to (iii). ∎

The preceding Theorem allows to express the optimal values for "larger" problems in terms of optimal values for "smaller" problems, but one needs to formalize the notation of "small" and "large" Since $G$ is acyclic, elements of $V$ can be topologically ordered so that any $x \in V$ is assigned an integer value $\pi(x)$ in the range of 1 to $n$.

Theorem 2.3 *The maximum weighted chain in a strongly pseudo-transitive graph $G = (V, E_1, E)$ can be computed in*

$O(\sum_{x \in V} deg^2(x) + n^2)$ *time, where $deg(x)$ denotes the degree of $x \in V$.*

Proof. Our algorithm is a dynamic programming algorithm, which uses the recurrence relations in Theorem 2.2. We will use both the adjacency matrix and also the adjacency list for the graph *G*. We first obtain a topological ordering of elements of *V* and compute $\pi(x)$ for any $x \in V$ in $O(n+m)$ time. We first obtain a topological ordering of elements of *V* and compute $\pi(x)$ for any $x \in V$ in $O(n+m)$ time. Initially, we will obtain a sorted list of edges $xy \in E$ in the increasing order of $\pi(y) - \pi(x)$.

We use four $n \times n$ *matrices* to store $\omega_{x,y}, \omega^1_{x,y}, \bar{\omega}_{x,y}$, and $\bar{\omega}^1_{x,y}$ for all $xy \in E$. We only compute the entries $xy$ of these matrices for which $xy \in E$ in the increasing of $\pi(y) - \pi(x)$ using the sorted list. Note that for any $x \in V$ there are $deg(x)$ many $y$'s for which $\omega$'s must be computed. Computing the LHS of recurrence relations in Theorem 2.2 can be done in $O(deg(x))$ time any fixed $xy \in E$.

It follows that total time is $O(\sum_{x \in V} deg^2(x) + n^2)$. ∎

Acknowledgement: Preliminary versions of these results were presented at DIMCAS workshop in Geometric Graph Theory, Rutgers University, 2002.